\documentclass{appolb}
\usepackage{pstricks}
\usepackage[dvips,final]{graphicx}
\usepackage{amsmath,amssymb,bm,pifont}

\begin{document}

\title{\hfill \hbox{RUB-TPII-05/2010}\vspace*{5mm}\\
       Pion Distribution Amplitude and Photon-to-Pion Transition Form Factor in QCD
       \thanks{Presented by the the first author at the International Meeting
       ``Excited QCD'', January 31--February 5, 2010,
               Tatranska Lomnica (Slovakia)}}

\author{Alexander~P.~Bakulev and S.~V.~Mikhailov
 \address{Bogoliubov Laboratory of Theoretical Physics, JINR \\
          Dubna 141980, Russia\\
          E-mail: bakulev@theor.jinr.ru}
\and N.~G.~Stefanis
 \address{Institut f\"{u}r Theoretische Physik II,
          Ruhr-Universit\"{a}t Bochum\\
          D-44780 Bochum, Germany\\
          E-mail: stefanis@tp2.ruhr-uni-bochum.de}
}
\maketitle
\begin{abstract}
We discuss the status of the pion distribution amplitude (DA)
in connection with QCD sum rules and experimental data
on the $\gamma^*\gamma\to \pi^0$ transition form factor.
Contents: (a) Pion DA in generalized QCD Sum Rules (SRs);
(b) Light Cone Sum Rules (LCSR) analysis of the CLEO data
for the $\gamma^*\gamma\to\pi^{0}$ transition form factor;
(c) Recent lattice QCD data for the pion DA;
(d) BaBar data---a challenge for QCD?
\end{abstract}

\markboth{\large \sl \underline{A.~P.~Bakulev et al.}
          \hspace*{2cm} Excited QCD, 2010}
         {\large \sl \hspace*{1cm} Pion DA and $\gamma^*\gamma\to\pi$ FF in QCD}

\noindent\textbf{Pion distribution amplitude from QCD sum rules}\\
The twist-two pion distribution amplitude (DA) parameterizes
the matrix element of the nonlocal axial current on the light cone
\cite{Rad77}
\begin{eqnarray}
\label{eq:pion.DA.ME}
  \langle{0\!\mid\!\bar d(z)\gamma_{\mu}\gamma_5
  [z,0]u(0)\!\mid\!\pi(P)}\rangle\Big|_{z^2=0}
= i f_{\pi}P_{\mu}\!
       \int\limits_{0}^{1}\! dx\ e^{ix(z\cdot P)}
        \varphi_{\pi}^\text{Tw-2}(x,\mu^2)\,.~~
\end{eqnarray}
The gauge-invariance of this DA is ensured by
the lightlike gauge link $[z,0]$,
inserted between the two separated quark fields.
The physical meaning of this DA is quite evident:
it is the amplitude
for the transition
$\pi(P)\rightarrow u(Px) + \bar{d}(P(1-x))$.
It is convenient to represent the pion DA using
an expansion
in terms of the Gegenbauer polynomials $C^{3/2}_n(2x-1)$,
which are the one-loop eigenfunctions of the ERBL
kernel~\cite{ER80,LB79}.
This representation means
that the whole scale dependence in
$\varphi_\pi(x;\mu^2)$ is transformed into
the scale dependence
of the Gegenbauer-coefficients
$a_2(\mu^2), a_4(\mu^2), \ldots$.

In order to construct reliable QCD SRs
for the pion DA moments,
one has to take into account the nonlocality of the QCD vacuum
condensates---as it has been shown in~\cite{MR86,BM98}.
As a concrete example for the nonlocal condensate (NLC) model,
we use here the minimal Gaussian model
$\displaystyle\langle{\bar{q}(0)q(z)}\rangle = \langle{\bar{q}\,q}\rangle\, e^{-|z^2|\lambda_q^2/8}$
with a single scale parameter $\lambda_q^2 = \langle{k^2}\rangle$,
characterizing
the average momentum of quarks in the QCD vacuum.
The value of $\lambda_q^2$ has been estimated in the
QCD SR approach and also on the lattice \cite{BI82lam,OPiv88,DDM99,BM02}:
$\lambda_q^2 = 0.35-0.55~\text{GeV}^2$.

The NLC SRs for the (twist-2) pion DA produce a bunch of
self-consistent two-parameter models at
the normalization scale $\mu_0^2\simeq 1.35$ GeV$^2$:
\begin{figure}[b!]
 \centerline{\includegraphics[width=0.48\textwidth]{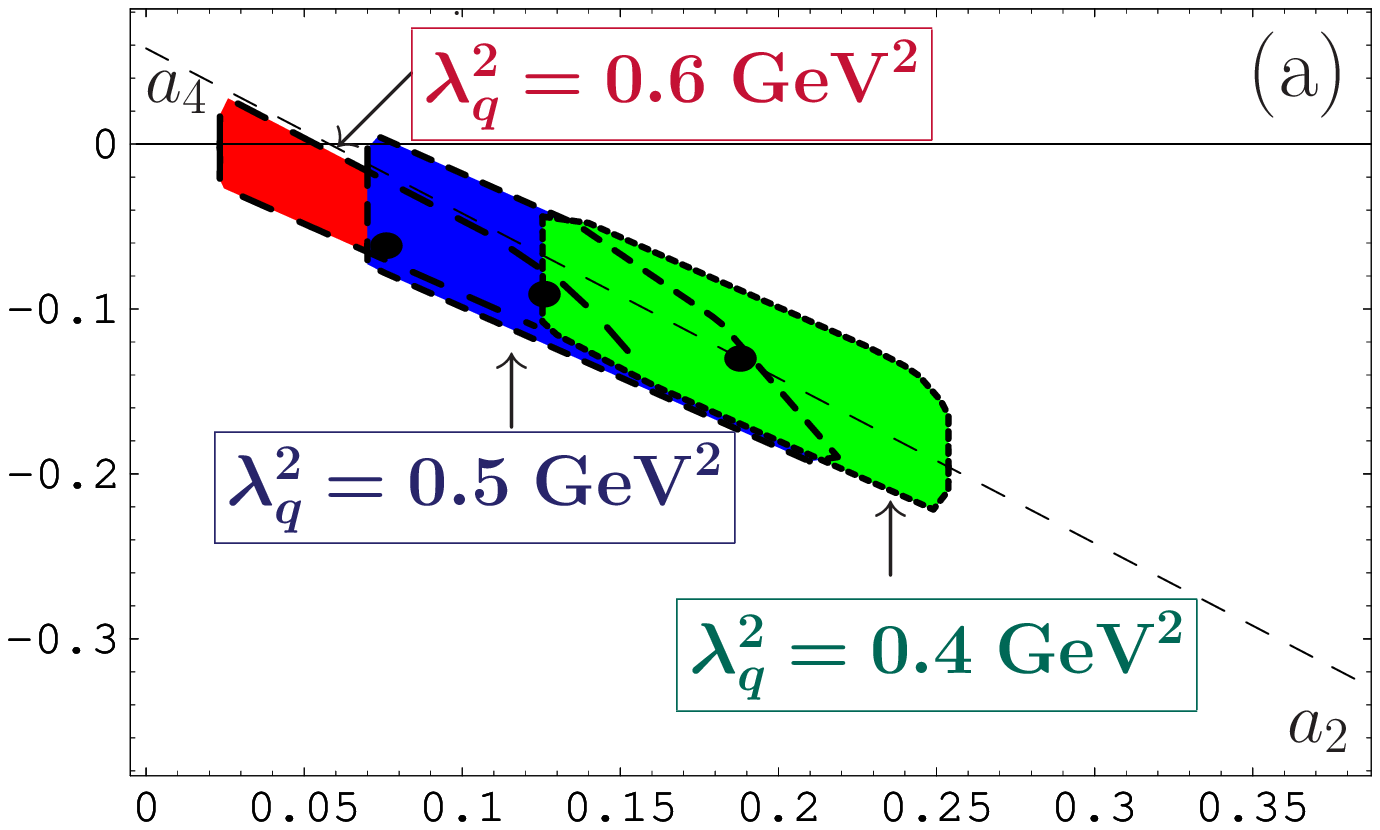}~~~
             \includegraphics[width=0.48\textwidth]{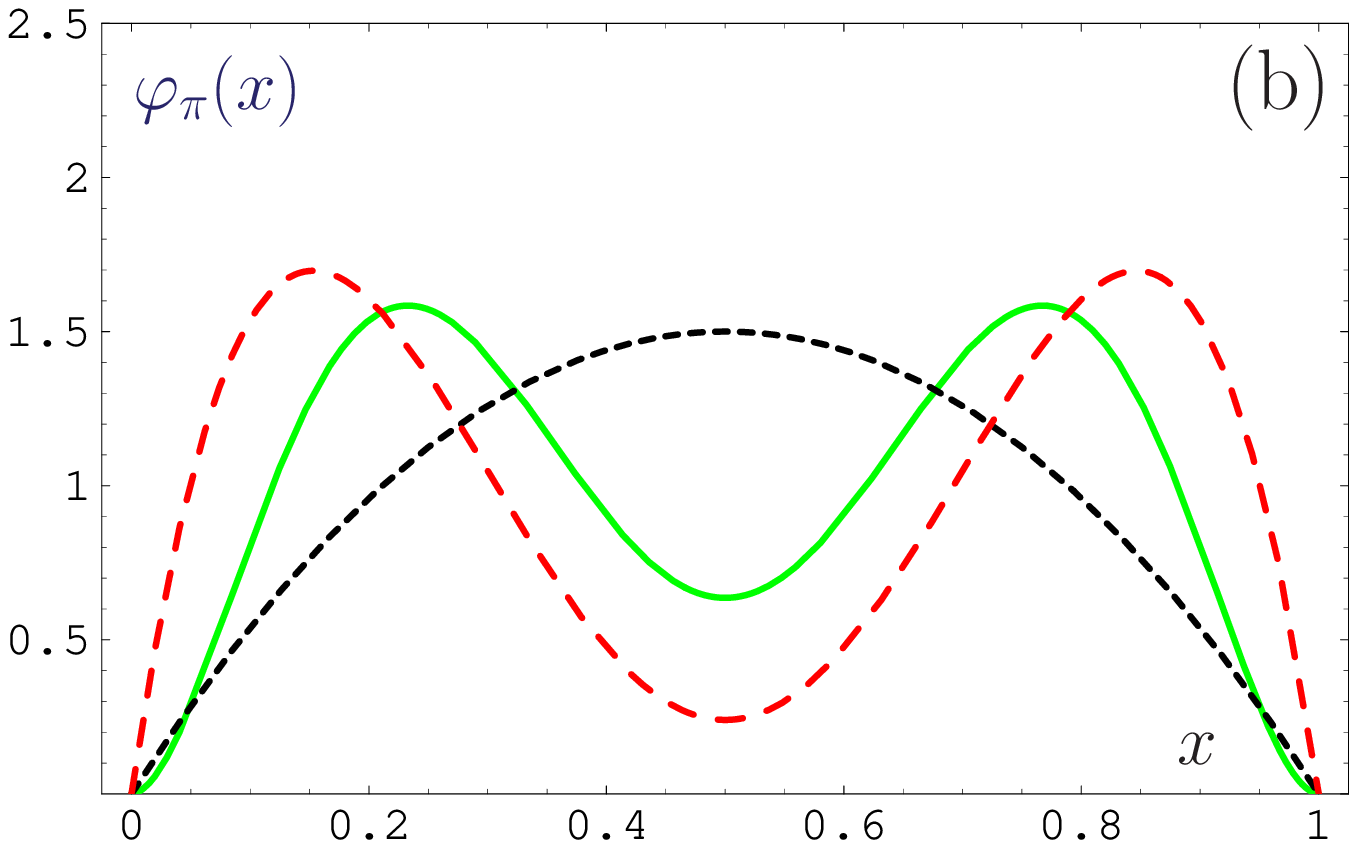}}
   \caption{\label{fig:456}
    \textbf{(a)}: The allowed values of the parameters
    $a_2$ and $a_4$
    of the pion DA bunch (\ref{eq:pi.DA.2Geg}),
    evaluated at $\mu^2=1.35$~GeV$^2$ for three values
    of the nonlocality parameter $\lambda_q^2=0.4\,, 0.5$,
    and $0.6$~GeV$^2$.
    \textbf{(b)}: Shapes of three characteristic pion
    DAs---BMS (solid line), CZ (dashed line),
    and the asymptotic DA (dotted line).}
\end{figure}
\begin{eqnarray}
 \label{eq:pi.DA.2Geg}
  \varphi^\text{NLC}_\pi(x;\mu_0^2)
  = \varphi^\text{As}(x)\,
     \Bigl[1 + a_{2}(\mu_0^2)\,C^{3/2}_{2}(2x-1)
             + a_{4}(\mu_0^2)\,C^{3/2}_{4}(2x-1)
     \Bigr]\,.~
\end{eqnarray}
The central point corresponds to
$a_2^\text{BMS}=+ 0.188$, $a_4^\text{BMS}=-0.130$
for $\lambda^2_q=0.4$ GeV$^2$,
whereas other allowed values of the parameters
$a_2$ and $a_4$---in
correspondence with associated values of
$\lambda^2_q$---are
shown in the left panel of Fig.\ \ref{fig:456}
in the form of slanted rectangles \cite{BMS01}.
All these solutions yield,
in accord with (\ref{eq:pi.DA.2Geg}),
the same value of the inverse moment of the pion DA, namely,
$\langle{x^{-1}}\rangle^\text{bunch}_{\pi} = 3.17\pm0.20$.
This range is in good agreement with
the estimates derived from a dedicated
SR for this moment
which can be obtained
through the basic SR
by integrating over $x$ with the corresponding weight
$x^{-1}$ (at $\mu_0^2\simeq 1.35$ GeV$^2$):
$\langle{x^{-1}}\rangle_{\pi}^{\text{SR}}=3.30\pm0.30$.
It is worth emphasizing at this point
that the moment
$\langle{x^{-1}}\rangle^\text{SR}_{\pi}$ can be safely determined
only
with the use of NLC SRs
because of the absence of endpoint singularities.

Comparing the obtained pion DA with the Chernyak--Zhitnitsky
(CZ) one \cite{CZ82}, reveals that, although both DAs are two-humped,
they are quite different, with the BMS DA being strongly endpoint
suppressed (see Fig.\ \ref{fig:456}(a)).

Qualitatively similar results have been also obtained with
the improved Gaussian model of the nonlocal QCD vacuum \cite{BP06}.
In that case, the allowed region for the parameters
$a_2$ and $a_4$ in Fig.\ \ref{fig:456}(a)
is shifted along the diagonal
farther to the right
with the central point being located near
the right corner of the allowed (green) region.
We emphasize that the BMS model \cite{BMS01}
is inside the allowed region
obtained with the improved QCD vacuum model.
This means that the characteristic features of the BMS bunch
persist for the improved bunch as well---in particular the
endpoint suppression.

\vspace*{+3mm}\noindent\textbf{Analysis of the CLEO data
 on $F_{\gamma\gamma^*\pi}(Q^2)$ and the pion DA}\\
\begin{figure}[b!]
 \centerline{\begin{minipage}{\textwidth}
  \centerline{\includegraphics[width=0.49\textwidth]{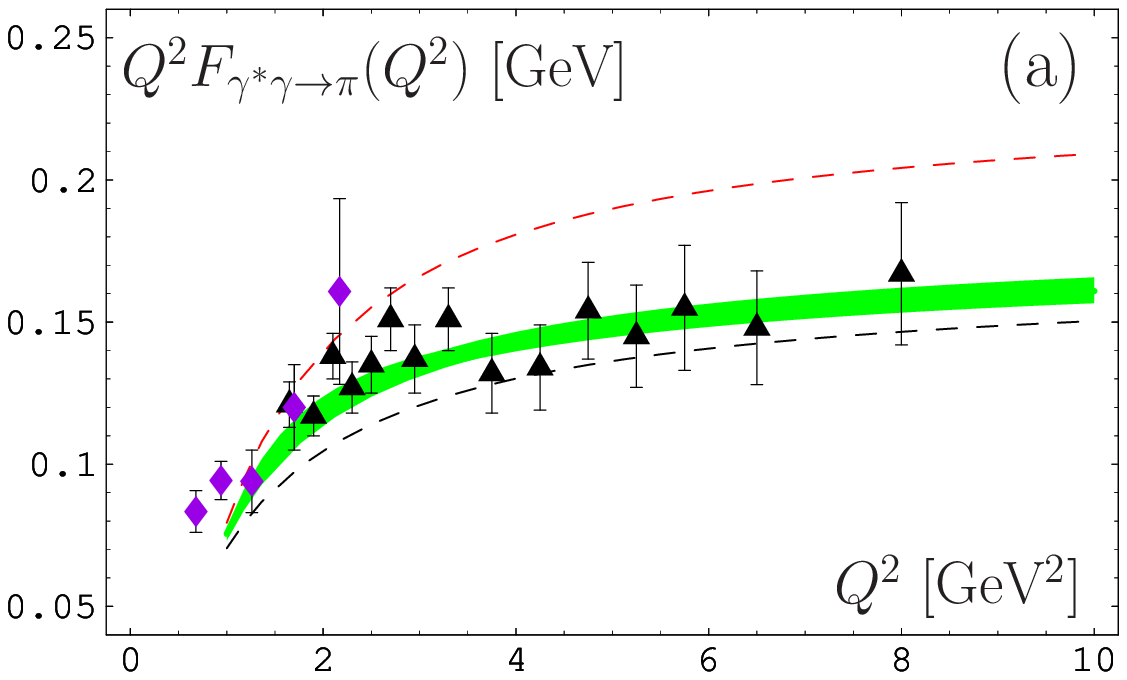}%
             ~\includegraphics[width=0.49\textwidth]{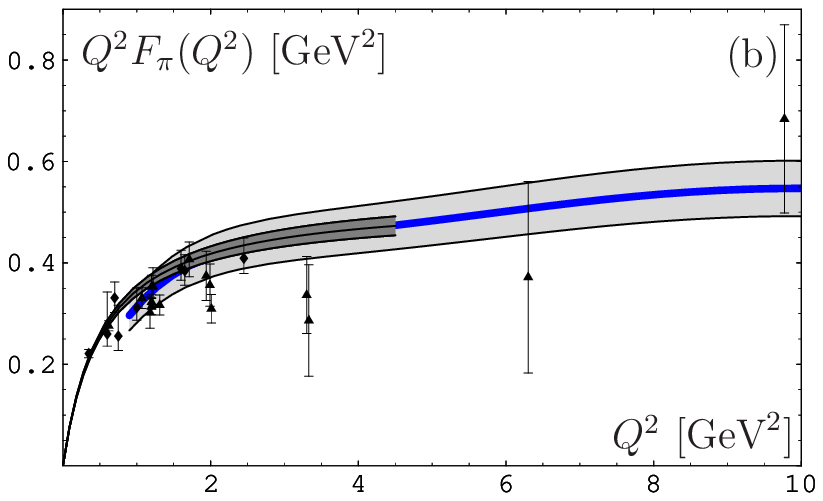}}
   \end{minipage}}
  \caption{\textbf{(a)}: LCSR predictions for
    $Q^2F_{\gamma^*\gamma\to\pi}(Q^2)$
    for the CZ DA (upper dashed line),
    BMS-``bunch'' (shaded strip),
    and the asymptotic DA (lower dashed line)
    in comparison with the CELLO
   (diamonds, \protect\cite{CELLO91}) and the CLEO
   (triangles, \protect\cite{CLEO98}) experimental data,
   evaluated with the twist-4 parameter value
   $\delta_{\rm Tw-4}^2=0.19$~GeV$^2$~\protect\cite{BMS02}
   and at $\mu^2_\text{SY}=5.76~\text{GeV}^2$.
   \textbf{(b)}: The scaled pion form factor
    calculated in the NLC QCD SRs (solid blue line)
    including nonperturbative uncertainties (shaded strip)~\protect{\cite{BPS09}}.
    The dark gray strip shows the lattice data~\cite{Brommel06}
    and the experimental data are taken from \protect{\cite{JLAB00}}
    (diamonds) and \cite{FFPI73}, \cite{FFPI76} (triangles).
    \label{fig:cello.cebaf}}
\end{figure}
Many studies \cite{SY99,SSK99,SSK00,AriBro-02,BM02,BMS02,Ag05a} have been
performed to determine the pion DA characteristics
using the high-precision CLEO data \cite{CLEO98}
on the pion-photon transition
form factor
$F_{\pi\gamma^{*}\gamma}(Q^2)$.
For instance, we have used in \cite{BMS02} LCSRs \cite{Kho99,SY99}
with next-to-leading-order accuracy
of QCD perturbation theory
with the aim to analyze the theoretical uncertainties
involved in the CLEO-data analysis
and extract more reliable estimates for the first two coefficients
$a_2$ and $a_4$,
which parameterize the deviation
from the asymptotic expression $\varphi_{\pi}^\text{As}$.
The upshot of our analysis \cite{BMS02,BMS03}
is that the CZ DA is excluded
at the $4\sigma$-level,
whereas the asymptotic DA is off at the $3\sigma$-level,
while the BMS DA (and most of the BMS bunch)
is inside the 1$\sigma$-error ellipse,
and the instanton-based model of Ref.\ \cite{PR01}
is close to the 2$\sigma$-boundary.
Moreover, we found that the CLEO data conform
with the value of the QCD nonlocality parameter
$\lambda_q^2 = 0.4$ GeV$^2$.

\vspace*{+3mm}\noindent\textbf{Pion form factor and JLab data}\\
In this context it is worth mentioning
our analysis of
the pion's electromagnetic form factor
which employs NLC QCD SRs,
Analytic QCD Perturbation Theory,
and NLC-derived pion DAs \cite{BPS09}.
The obtained results
are in excellent agreement with
the recent JLab data
and also with the lattice data of Ref.\ \cite{Brommel06},
as one sees from Fig.\ \ref{fig:cello.cebaf}(b),
where the light gray strip includes the NLC QCD SRs uncertainties,
whereas the dark gray strip represents the lattice results.

\vspace*{+3mm}\noindent\textbf{New lattice data and the pion DA}\\
\begin{figure}[b!]
 \centerline{\begin{minipage}{0.49\textwidth}
             \includegraphics[width=\textwidth]{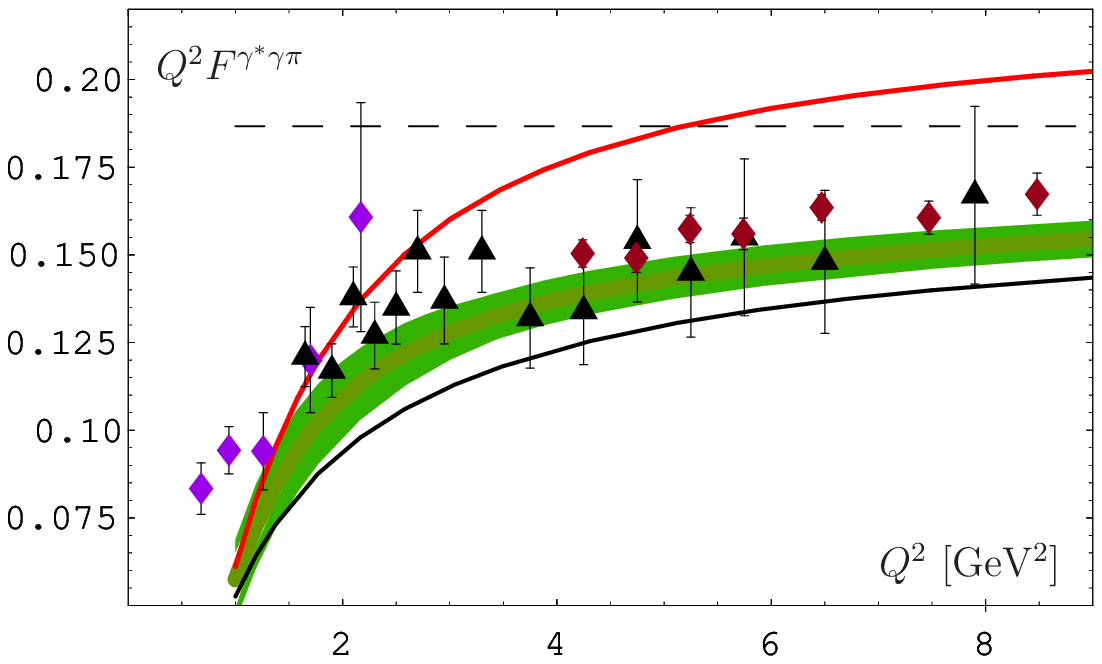}\end{minipage}
          ~~~\begin{minipage}{0.46\textwidth}\vspace*{-1mm}
           \includegraphics[width=\textwidth]{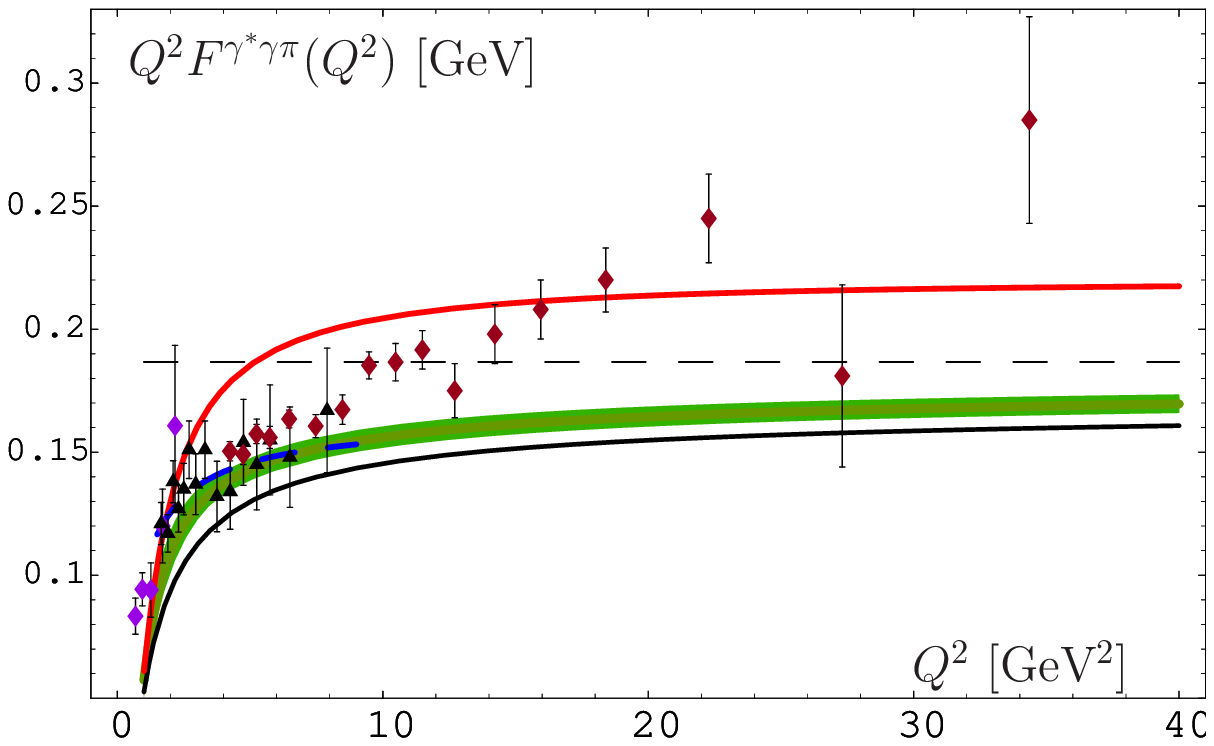}\end{minipage}}
  \caption{NNLO$_{\beta_0}$ LCSR predictions for
    $Q^2F_{\gamma^*\gamma\to\pi}(Q^2)$
    for the CZ DA (upper red line),
    BMS-``bunch'' (shaded green strip),
    and the asymptotic DA (lower solid line)
    in comparison with the CELLO, CLEO (the same as in Fig.\ \ref{fig:456}(a))
    and BaBar~\cite{BaBar09} (brown diamonds with much smaller error-bars)
    experimental data.
    \textbf{Left panel}: $Q^2\leq9$~GeV$^2$ region.
    \textbf{Right panel}: whole BaBar region of $Q^2$.
    \label{fig:Babar}}
\end{figure}
Rather recently,
new high-precision lattice measurements
of the second moment of the pion's DA
$\langle{\xi^2}\rangle_{\pi} = \int_0^1(2x-1)^2\varphi_\pi(x)\,dx$
appeared \cite{DelD05,Lat06}.
Both groups extracted from their respective simulations
values of $a_2$ at the Schmedding--Yakovlev scale
$\mu^2_\text{SY}\approx 0.24$,
but with different error bars.
Remarkably,
these lattice results are in striking agreement
with the estimates for $a_2$ both from NLC QCD SRs~\cite{BMS01}
and also from the CLEO-data analyses
based on LCSRs \cite{SY99,BMS02}.
The improved bunch \cite{BP06} appears to have an even better agreement
with the recent lattice results of \cite{Lat06}.


\vspace*{+1mm}\noindent\textbf{Confronting NNLO LCSR results for
$F_{\gamma^*\gamma\to\pi}$ with the BaBar data}\\
In a recent paper by two of us \cite{MS09},
the NNLO QCD radiative corrections,
proportional to the $\beta_0$-coefficient,
have been calculated and included into the LCSR analysis
of the pion-photon transition form factor.
The overall effect of these corrections appears to be negative,
hence reducing the $\gamma^*\gamma\to\pi^0$ form factor
by $-7$\% at low $Q^2\sim2$~GeV$^2$ and
by $-2.5$\% at intermediate $Q^2\gtrsim6$~GeV$^2$
(left panel in Fig.\ \ref{fig:Babar}).
In spite of this reduction, the BMS bunch describes rather well all data
for $Q^2\in[1.5,9]$~GeV$^2$---including those of BaBar \cite{BaBar09}.
This means that the CLEO (and the low-energy BaBar) data
are incompatible
with wide pion DAs and demand
that the endpoints $x=0,1$ are stronger suppressed
than in the asymptotic DA.

Surprisingly, the BaBar data~\cite{BaBar09}
contradict this behavior in the high-energy region:
there they show a significant growth
with $Q^2$
for values above $\sim 10$~GeV$^2$.
This behavior of the BaBar data is clearly in conflict
with the collinear
factorization in perturbative QCD.
This is true for any pion DA that vanishes at the endpoints $x=0,1$
(see Fig.\ \ref{fig:Babar}---right panel)
and the explanations in \cite{MS09}.

Despite the appearance of several proposals to explain
the anomalous behavior of the high-$Q^2$ BaBar data \cite{Dor09,Rad09,Pol09,Che09},
we want to emphasize
that from the point of view of QCD
it is not possible to describe them
either with the inclusion of higher radiative corrections
or with higher-twists---both types of contributions
are negative \cite{BMS02,MS09,BMS05lat}.

\vspace*{+3mm}\noindent\textbf{Conclusions:}\\
To conclude, the QCD SRs produce an endpoint-suppressed bunch of pion DAs
in agreement with the CELLO and the CLEO data on
$F_{\gamma^*\gamma\to\pi}$.
The values of the second moment of these DAs are within the range
found in recent lattice simulations.
These results are also in line with the new BaBar data up to
$\sim 10$~GeV$^2$
but do not reproduce
the observed growth above this scale.
It was shown in \cite{MS09Trento} that the CELLO and the CLEO data
cannot be fitted simultaneously with all the BaBar data.

\vspace*{+3mm}\noindent\textbf{Acknowledgments}\\
This investigation was supported in part
by the Heisenberg--Landau Program, Grant 2010,
and the Russian Foundation for Fundamental Research,
Grants No.\ 07-02-91557, 08-01-00686 and 09-02-01149.



\end{document}